\documentclass[manuscript]{aastex}

\title{Abundance Anomaly of the $^{13}$C Isotopic Species of c-C$_3$H$_2$ in the Low-Mass Star Formation Region L1527\altaffilmark{\dag}}
\slugcomment{Ver. \today}
\author{	Kento Yoshida\altaffilmark{1}, 
		Nami Sakai\altaffilmark{1}, 
		Tomoya Tokudome\altaffilmark{1}, 
		Ana L\'opez-Sepulcre\altaffilmark{1}, 
		Yoshimasa Watanabe\altaffilmark{1}, 
		Shuro Takano\altaffilmark{2,3}, 
		Bertrand Lefloch\altaffilmark{4,5}, 
		Cecilia Ceccarelli\altaffilmark{4,5}, 
		Rafael Bachiller\altaffilmark{6}, 
		Emmanuel Caux\altaffilmark{7,8}, 
		Charlotte Vastel\altaffilmark{7,8}, 
		and Satoshi Yamamoto\altaffilmark{1}}
\altaffiltext{\dag}{Based on observations carried out with the IRAM 30 m Telescope and the NRO 45 m Telescope. IRAM
is supported by INSU/CNRS (France), MPG (Germany) and IGN (Spain). NRO is a branch of National Astronomical Observatory of Japan, National Institutes of Natural Sciences, Japan.}
\altaffiltext{1}{Department of Physics, The University of Tokyo, Bunkyo-ku, Tokyo 113-0033, Japan}
\altaffiltext{2}{Nobayama Radio Observatory, Minamimaki, Minamisaku, Nagano 384-1305, Japan}
\altaffiltext{3}{Department of Astronomical Science, The Graduate University for Advanced Studies (SOKENDAI), 2-21-1 Osawa, Mitaka, Tokyo 181-8588, Japan}
\altaffiltext{4}{Universit\'e de Grenoble Alpes, IPAG, F-38000 Grenoble, France}
\altaffiltext{5}{CNRS, IPAG, F-38000 Grenoble, France}
\altaffiltext{6}{Observatorio Astron\'omico Nacional (OAN, IGN). Calle Alfonso XII 3, E-28014 Madrid, Spain}
\altaffiltext{7}{Universit\'e de Toulouse, UPS-OMP, IRAP, Toulouse, France}
\altaffiltext{8}{CNRS, IRAP, 9 Av. Colonel Roche, BP 44346, F-31028 Toulouse Cedex 4, France}
\begin{document}
\begin{abstract}
The rotational spectral lines of c-C$_3$H$_2$ and two kinds of the $^{13}$C isotopic species, c-$\mathrm{^{13}CCCH_2}$ ($C_{2v}$ symmetry) and c-$\mathrm{CC^{13}CH_2}$ ($C_s$ symmetry) have been observed in the 1--3 mm band toward the low-mass star-forming region L1527. We have detected 7, 3, and 6 lines of c-C$_3$H$_2$, c-$\mathrm{^{13}CCCH_2}$, and c-$\mathrm{CC^{13}CH_2}$, respectively, with the Nobeyama 45 m telescope, and 34, 6, and 13 lines, respectively, with the IRAM 30 m telescope, where 7, 2, and 2 transitions, respectively, are observed with the both telescopes. With these data, we have evaluated the column densities of the normal and $^{13}$C isotopic species. The [c-C$_3$H$_2$]/[c-$\mathrm{^{13}CCCH_2}$] ratio is determined to be $310\pm80$, while the [c-C$_3$H$_2$]/[c-$\mathrm{CC^{13}CH_2}$] ratio is determined to be $61\pm11$.
The [c-C$_3$H$_2$]/[c-$\mathrm{^{13}CCCH_2}$] and [c-C$_3$H$_2$]/[c-$\mathrm{CC^{13}CH_2}$] ratios expected from the elemental $^{12}$C/$^{13}$C ratio are 60--70 and 30--35, respectively, where the latter takes into account the statistical factor of 2 for the two equivalent carbon atoms in c-C$_3$H$_2$. Hence, 
this observation further confirms the dilution of the $^{13}$C species in carbon-chain molecules and their related molecules, which are thought to originate from the dilution of $^{13}$C$^+$ in the gas-phase C$^+$ due to the isotope exchange reaction: $^{13}\mathrm{C}^++\mathrm{CO}\rightarrow{}^{13}\mathrm{CO}+\mathrm{C}^+$.
Moreover, the abundances of the two $^{13}$C isotopic species are different from each other. The ratio of c-$\mathrm{^{13}CCCH_2}$ species relative to c-$\mathrm{CC^{13}CH_2}$ is determined to be $0.20\pm0.05$. If $^{13}$C were randomly substituted for the three carbon atoms, the [c-$\mathrm{^{13}CCCH_2}$]/[c-$\mathrm{CC^{13}CH_2}$] ratio would be 0.5. Hence, the observed ratio indicates that c-$\mathrm{CC^{13}CH_2}$ exists more favorably.
Possible origins of the different abundances are discussed.
\end{abstract}

\section{Introduction}
Radio spectral lines of $^{13}$C isotopic species of various molecules are now detectable with a reasonable observation time, thanks to recent developments of receiver and backend technologies. 
So far, spectral lines of $^{13}$C isotopic species of abundant molecules, whose main isotopologue  (normal species) lines are optically thick, have widely been observed to derive the column density of the normal species accurately. 
In this case, the $^{12}$C/$^{13}$C ratio of each species is usually assumed to be the same as the elemental $^{12}$C/$^{13}$C ratio, which is 60--70 in the solar neighborhood \citep{lucas, milam}. However, anomalies of the $^{12}$C/$^{13}$C ratio for carbon-chain molecules have recently been recognized in nearby cold and dense molecular clouds. The anomalies include different $^{12}$C/$^{13}$C ratios for different carbon atoms in a single molecular species and significantly high $^{12}$C/$^{13}$C ratios for several carbon-chain molecules in comparison with the elemental $^{12}$C/$^{13}$C ratio.

As for the former anomaly, \citet{takano} reported the non-equivalent abundances of the three $^{13}$C species of HC$_3$N toward the cold dark cloud TMC-1 (Cyanopolyyne peak; CP). The abundance of the HCC$^{13}$CN is higher than that of H$^{13}$CCCN and HC$^{13}$CCN by a factor of 1.4. \citet{s07} also reported a large difference between the $^{13}$CCS and C$^{13}$CS abundances toward TMC-1 (CP), the ratio of C$^{13}$CS/$^{13}$CCS being 4.2. 
A similar non-equivalence of the $^{13}$C isotopologues of a single molecular species is reported for CCH, C$_3$S, and C$_4$H \citep{s10a,s13}. 
It is proposed that these non-equivalent abundances of the $^{13}$C species reflect production pathways of each molecule and/or originate from gas phase reactions interchanging the $^{13}$C position within a molecule \citep{s10a,furuya}.

The second anomaly is that the $^{12}$C/$^{13}$C ratios of some carbon-chain molecules are significantly higher than the elemental $^{12}$C/$^{13}$C ratio (60--70) in the solar neighborhood, indicating that the $^{13}$C isotope is diluted in these molecules. The most striking result is reported for CCH \citep{s10a}. The CCH/$^{13}$CCH and CCH/C$^{13}$CH ratios are higher than 250 and 170 toward TMC-1 (CP), respectively. High $^{12}$C/$^{13}$C ratios are also seen in CCS, C$_3$S, and C$_4$H. Such a dilution of the $^{13}$C isotopic species was theoretically predicted by \citet{langer}, and it was confirmed for some carbon-chain molecules by the above observations. On the other hand, \citet{liszt} reported the $^{12}$C/$^{13}$C ratios of CS, HNC, and H$_2$CS to be in the range from 43 to 79 toward TMC-1 (CP), which is close to the elemental $^{12}$C/$^{13}$C ratio. Hence, the degree of the $^{13}$C dilution seems dependent on molecular species, although its origin is still controversial.

It is therefore important to explore whether the $^{13}$C isotope anomalies are seen in other molecules. Furthermore, the above observations were mostly carried out toward TMC-1 (CP), although the anomaly was reported  for CCS toward the cold dark cloud L1521E and for CCH toward the low-mass star-forming region L1527. With these in mind, we here report on the observations of the $^{13}$C isotopic species of c-C$_3$H$_2$ toward the low-mass star-forming region L1527. 

c-C$_3$H$_2$ is a carbon-chain related molecule, and is known to exist widely in interstellar clouds \citep[e.g.,][]{madden89, cox}.
It has two kinds of $^{13}$C isotopic species, as shown in Figure \ref{structure}. The off-axis $^{13}$C species (referred to hereafter as c-$\mathrm{CC^{13}CH_2}$) is statistically favored by a factor of 2, because there are two equivalent carbon atoms in c-C$_3$H$_2$ for $^{13}$C substitution for this species. It was detected in TMC-1 and other sources \citep{madden86, gomez, spezzano}. On the other hand, detection of the on-axis $^{13}$C species (referred to hereafter as c-$\mathrm{{}^{13}CCCH_2}$) has not been reported in cold clouds as far as we know, although detection of one line was claimed by \citet{gomez} toward the Galactic center cloud Sgr B2. During the spectral line surveys of L1527 with Nobeyama 45 m telescope (a legacy program of Nobeyama Radio Observatory) and IRAM 30 m telescope (the IRAM large program ASAI), we detected many spectral lines of the off-axis and on-axis $^{13}$C isotopic species as well as those of the main isotopologue. On the basis of these extensive datasets, we have explored the $^{13}$C abundance anomaly of c-C$_3$H$_2$ in L1527.

\section{Observation}

\subsection{Observation with Nobeyama 45 m Telescope}
The spectral line survey of the 3 mm band (80 GHz to 116 GHz) was carried out with Nobeyama 45 m telescope (referred to hereafter as NRO 45 m) toward L1527 from 2006 to 2011, as the legacy project of the Nobeyama Radio Observatory. 
Transition frequencies and quantum numbers of the observed lines for c-$\mathrm{C_3H_2}$ and its $\mathrm{{}^{13}C}$ species are listed in Tables 1--3.
The observed position was $(\alpha_{2000},\ \delta_{2000})=(04^\mathrm{h}39^\mathrm{m}53\fs89,\ 26\arcdeg03\arcmin11\farcs0)$.
In this observation, we used  two SIS mixer receivers, S80 and S100, simultaneously in 2006--2008 and the dual-polarization sideband-separating SIS receiver T100H/V \citep{nakajima} in 2009--2011.
The system temperatures varied from 250 to 350 K in 2006--2008 and from 150 to 250 K in 2009--2011.
The pointing of the telescope was checked every $1\sim1.5$ hours by observing nearby SiO maser sources (NML-Tau, Ori-KL), and was maintained to be better than 6\arcsec.
The intensity scale was calibrated by the chopper-wheel method, and is estimated to be accurate  within 20\%.
The main-beam efficiency is 0.43 in 2006--2008, 0.49 in 2009, 0.42 in 2010--2011, as reported in the NRO website\footnote{http://www.nro.nao.ac.jp/\~{}nro45mrt/html/prop/status/Status\_R14.html}.
The beam size is 19\arcsec\  at 86 GHz\footnotemark[1].
In 2006--2011, we used a bank of acousto-optical radio spectrometers (AOSs) whose bandwidth, resolution, and channel spacing are 250 MHz, 250 kHz, and 125 kHz, respectively. In 2012, we used a bank of autocorrelators, SAM45, whose bandwidth, resolution, and channel spacing were set to be 500 MHz, 244 kHz, and 244 kHz, respectively. The frequency resolution corresponds to $\sim0.8$ km s$^{-1}$ at 90 GHz. 
This resolution is not sufficient, because the typical line width toward this source is from $0.5$ to 0.8 km s$^{-1}$. However, we chose this resolution to cover the whole 3 mm band within the limited observation time. Although the lines are partly frequency-diluted, the integrated intensity is reliable. A position switching mode with the off-position, $(\alpha_{2000},\ \delta_{2000})=(04^\mathrm{h}42^\mathrm{m}35\fs9,\ 25\arcdeg53\arcmin23\farcs3)$, was employed in all the observations.

\subsection{Observation with IRAM 30 m Telescope}
Observations with IRAM 30 m telescope (referred to hereafter as IRAM 30 m) at Pico Veleta were performed in several runs between 2012 and 2013, as part of the ASAI (Astrochemical Surveys at IRAM) Large Program. The survey covers the spectral bands at 3 mm (80--112 GHz), 2 mm (130--173 GHz), and 1 mm (200--276 GHz).
The beam size is 29\arcsec, 17\arcsec, and 12\arcsec\  at 86, 145, and 210 GHz, respectively, as reported in the IRAM website\footnote{http://www.iram.es/IRAMES/mainWiki/Iram30mEfficiencies}.
The wobbler switching mode was employed with a beam throw of 180\arcsec. This amount of the beam throw is enough, because most of the emission comes from the 15\arcsec--20\arcsec\ diameter area around the protostar according to the interferometer observations combined with single-dish observations \citep{s10b}. The frontends were the broad-band EMIR receivers, and the backends were the FTS spectrometers in their 195 kHz resolution mode, corresponding to velocity resolutions of 0.21--0.73 km s$^{-1}$.
The antenna temperature ($T_\mathrm{A}^\ast$) was converted to the main-beam brightness temperature ($T_\mathrm{mb}$) by dividing by $B_\mathrm{eff}/F_\mathrm{eff}$. Here, $B_\mathrm{eff}$ and $F_\mathrm{eff}$ are beam efficiency and forward efficiency, respectively. The $B_\mathrm{eff}/F_\mathrm{eff}$ value is 0.85, 0.80, and 0.67 at 86, 145, and 210 GHz, respectively\footnotemark[2].

\section{Results}
\subsection{Overall results}
\begin{figure}
\begin{center}
\includegraphics[width=\textwidth]{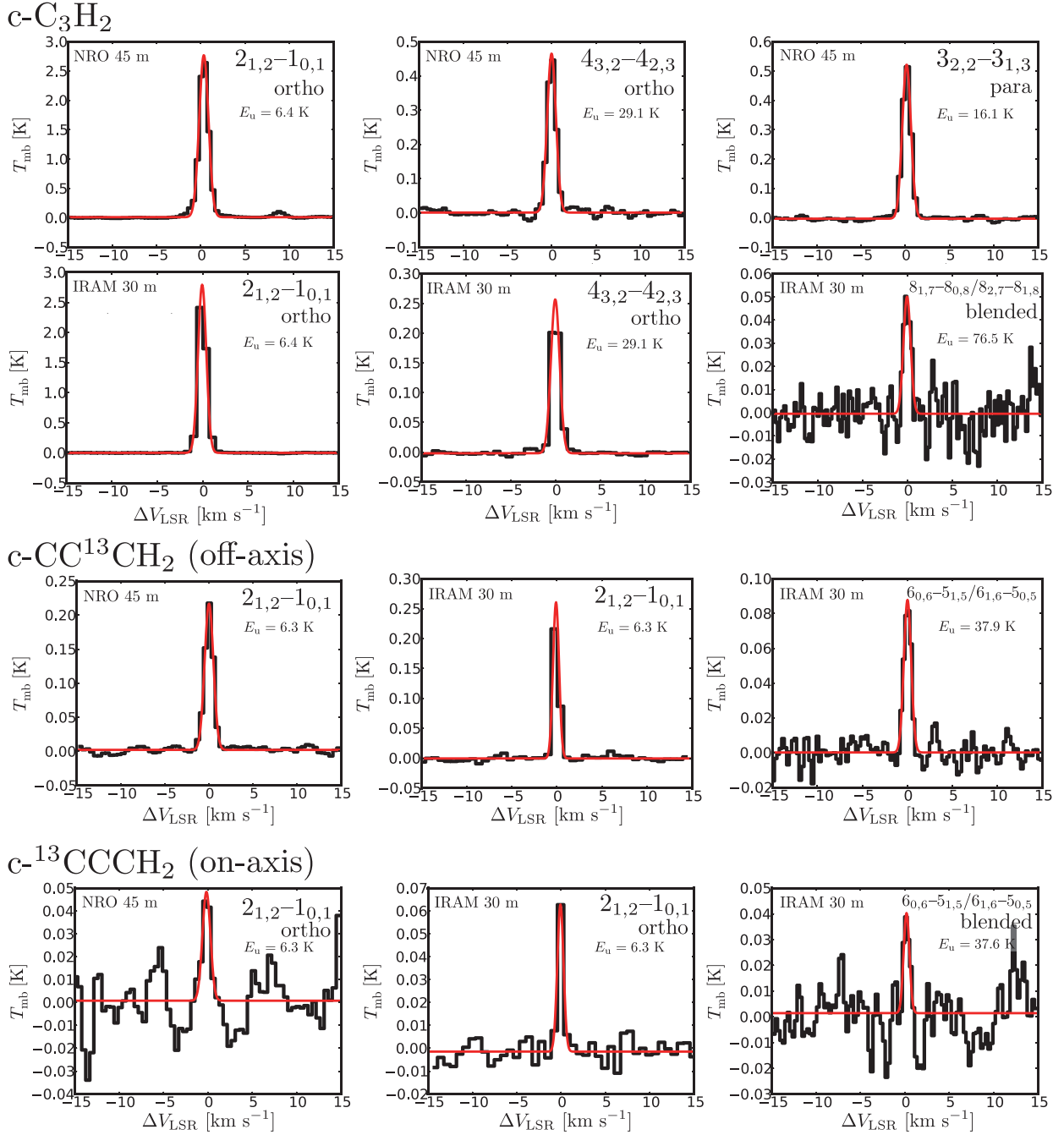}
\caption{Observed line profiles of c-C$_3$H$_2$ and its $^{13}$C species in L1527. The results of Gaussian fitting are also shown in red.}
\label{lines}
\end{center}
\end{figure}

\begin{figure}
\begin{center}
\includegraphics[width=\textwidth]{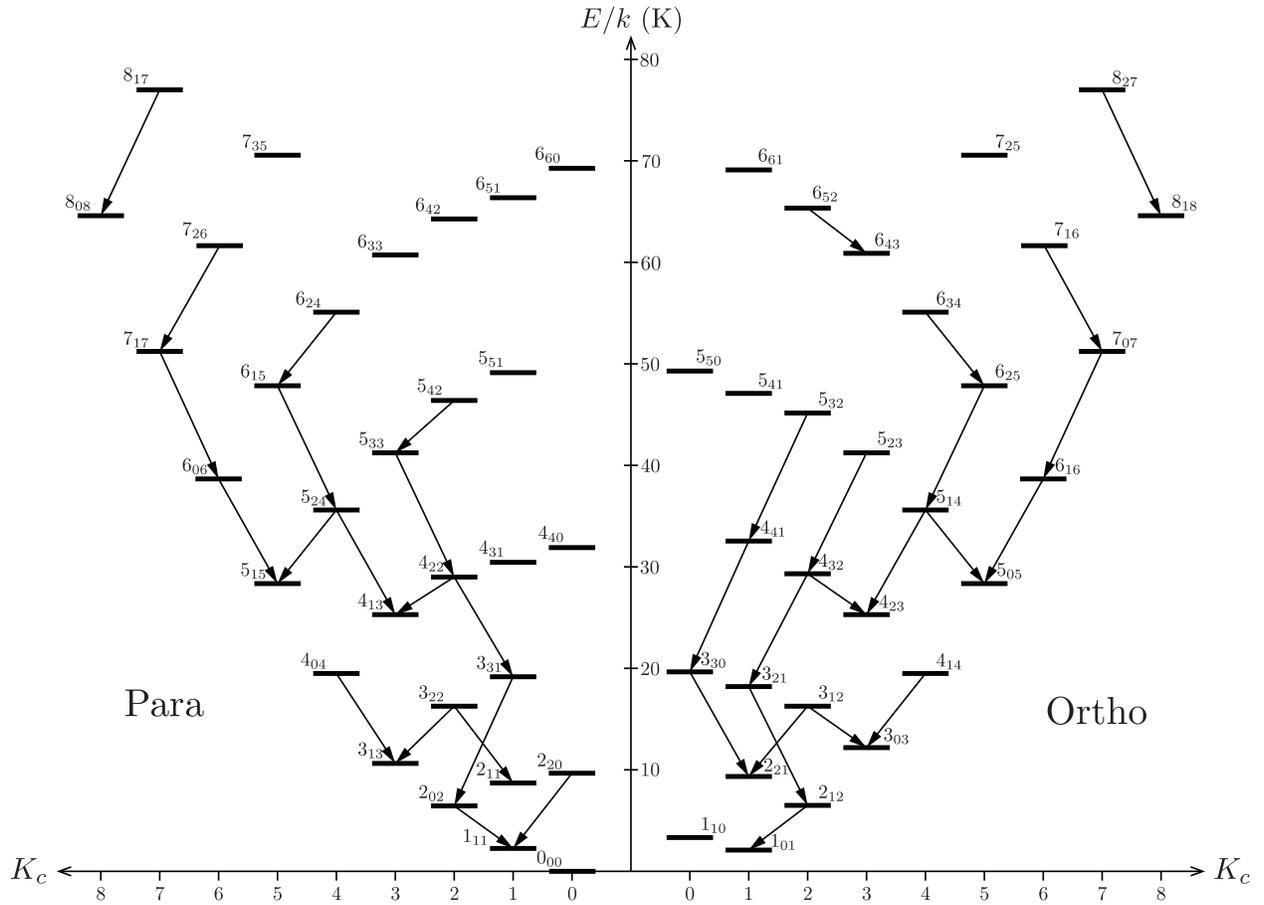}
\caption{The energy level diagram of c-C$_3$H$_2$. The transitions observed in L1527 are shown by downward arrows.}
\label{diagram}
\end{center}
\end{figure}

In the Nobeyama observations, we detected 7 lines of c-C$_3$H$_2$ in the 3 mm band (Table \ref{N_of_lines}). 
Later, we further detected 8, 9, and 17 lines in the 3 mm, 2 mm, and 1 mm bands in the ASAI observations. 
Examples of the observed spectra are shown in Figure \ref{lines}, along with the best Gaussian fit to their profiles.
In total, 41 lines of c-C$_3$H$_2$ were detected, among which 7 transitions were observed both in the Nobeyama and ASAI observations. 
Even high excitation lines of $8_{1,8}$--$8_{0,8}$ and $8_{2,7}$--$8_{1,8}$, whose upper state energies are as high as 77 K (53 cm$^{-1}$), were detected. 
The detected lines are summarized in Tables \ref{normal} and \ref{N_of_lines}, and are also shown in the energy level diagram (Figure \ref{diagram}). Since c-C$_3$H$_2$ has a pair of equivalent H nuclei, the rotational levels are classified into the ortho and para species. Radiative and collisional interconversion between the ortho and the para species is strongly forbidden, and hence, the ortho and the para species behave as different molecular species. In our observations, 19 and 18 lines are detected for the ortho and para species, respectively, while the remaining 4 lines detected in the 1 mm band are blended lines of the ortho and para species.

\begin{figure}
\begin{center}
\includegraphics[width=.9\textwidth]{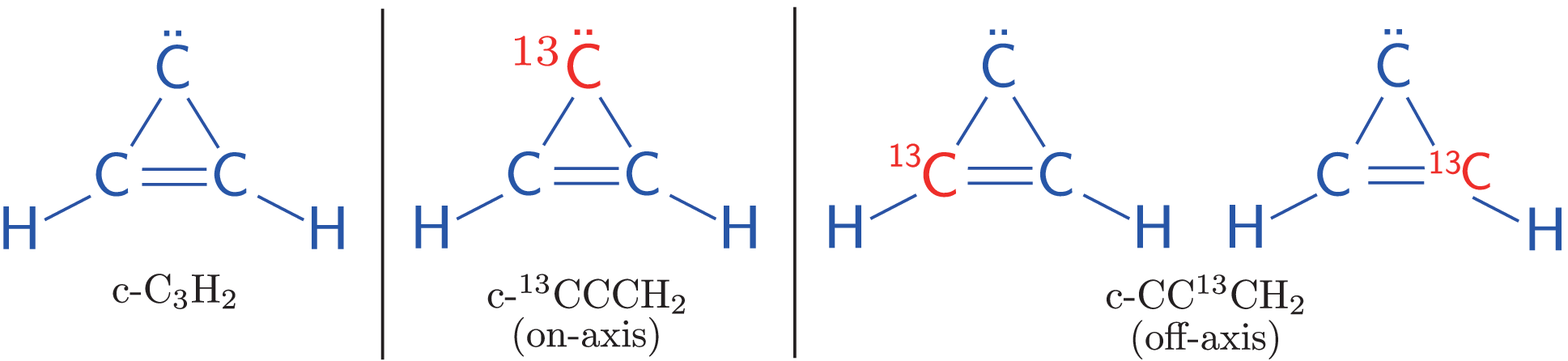}
\caption{Molecular geometry of c-C$_3$H$_2$ and its $^{13}$C species}
\label{structure}
\end{center}
\end{figure}

In addition to the lines of normal species, we also detected the lines of the $^{13}$C isotopic species. The line parameters of c-$\mathrm{CC^{13}CH_2}$ and c-$\mathrm{^{13}CCCH_2}$ species  (Figure \ref{structure}) are summarized in Tables \ref{off} and \ref{on}, respectively. For c-$\mathrm{CC^{13}CH_2}$, we detected 6 lines in the 3 mm band in the Nobeyama observations, and 2, 4, and 7 lines in the 3 mm, 2 mm, 1 mm bands, respectively, in the ASAI observations. In total, we detected 19 lines for c-$\mathrm{CC^{13}CH_2}$, as summarized in Tables \ref{off} and \ref{N_of_lines}. In contrast to the normal species, c-$\mathrm{CC^{13}CH_2}$ does not have the ortho and para classification, because the two hydrogen nuclei are no longer equivalent. For c-$\mathrm{^{13}CCCH_2}$, we detected 3 lines in the 3 mm band in the Nobeyama observations, and 2, 1, and 3 lines in the 3 mm, 2 mm, 1 mm bands, respectively, in the ASAI observations, as summarized in Tables \ref{on} and \ref{N_of_lines}. Hence, c-$\mathrm{^{13}CCCH_2}$ is definitively detected, as shown in Figure \ref{lines}. This species has the ortho and para states as in the case of the normal species. It should be noted that c-$\mathrm{CC^{13}CH_2}$ is statistically favored by a factor of 2 in comparison with c-$\mathrm{^{13}CCCH_2}$, because two equivalent carbon atoms are available in c-C$_3$H$_2$ for $^{13}$C substitution for c-CC${}^{13}$CH$_2$ while only one for c-${}^{13}$CCCH$_2$ (Figure \ref{structure}).

\subsection{LTE analysis for c-C$_3$H$_2$}
Assuming the local thermodynamic equilibrium (LTE) condition, we determine the rotation temperatures and the beam-averaged column densities of c-C$_3$H$_2$ and its $^{13}$C species by using the least-squares analysis with the following equations:
\[T_\mathrm{b} = \frac{h\nu}{k}\left[\frac{1}{\exp(h\nu/kT_\mathrm{rot})-1}-\frac{1}{\exp(h\nu/kT_\mathrm{bg})-1}\right](1-e^{-\tau}),\]
and
\[\tau=\frac{8\pi^3S\mu^2}{3h\Delta v U(T_\mathrm{rot})}\left[\exp\left(\frac{h\nu}{kT_\mathrm{rot}}\right)-1\right]\exp\left( -\frac{E_\mathrm{u}}{kT_\mathrm{rot}} \right)N,\]
where $T_\mathrm{b}$ is the brightness temperature, $h$ the Planck constant, $\nu$ the transition frequency, $k$ the Boltzmann constant, $T_\mathrm{rot}$ the rotation temperature, $T_\mathrm{bg}$ the cosmic microwave background temperature, $\tau$ the optical depth, $S$ the line strength, $\mu$ the dipole moment, $\Delta v$ the line width, $U(T_\mathrm{rot})$ the partition function, $E_\mathrm{u}$ the upper state energy, and $N$ the total column density.
The partition function is numerically calculated from energies and degeneracies of the rotational levels, which are taken from the spectroscopy database, CDMS \citep{muller} (See also original spectroscopy references: \citet{bogey86,vrtilek,bogey87,spezzano12}.).

Because the size of the emitting region is comparable to or smaller than the telescope beam size, the effect of beam dilution has to be taken into account. This correction for the beam dilution is particularly important in this analysis, because the 1, 2 and 3 mm data, which result in different beam sizes, are used simultaneously. Furthermore, the correction is also necessary for the 3 mm data taken with the two different telescopes. The correction for the beam dilution is made by dividing the observed intensity by the beam dilution factor
$\eta=\theta_\mathrm{s}^2/(\theta_\mathrm{s}^2+\theta_\mathrm{b}^2),$ 
where $\theta_\mathrm{s}$ and $\theta_\mathrm{b}$ denote the source diameter and the antenna beam width (FWHM), respectively. This correction factor assumes a Gaussian beam and a Gaussian distribution of the source emission.
The FWHM beam widths are assumed to be $1.7\times10^3/\nu$(GHz) and $2.4\times10^3/\nu$(GHz) in arcseconds for NRO 45 m and IRAM 30 m, respectively, on the basis of the beam sizes mentioned in Sections 2.1 and 2.2.
\citet{s10b} reported that c-$\mathrm{C_3H_2}$ is distributed within a diameter of 2000--3000 AU from the protostar, which corresponds to the size of 15\arcsec--20\arcsec.
Thus, the size of the emitting region is assumed to be 20\arcsec. 
The intensities of the high excitation lines ($E_\mathrm{u}\gtrsim30$ K) in the 3 mm band observed with NRO 45 m are brighter by a factor of 1.3--2.4 than those observed with IRAM 30 m, which further supports such a compact distribution.
In the least-squares analysis, the ortho and para species are treated as different molecules, and the ortho-to-para ratio is taken as a fitting parameter. This analysis is possible owing to the large dataset of the observed line intensities.
Relative weights of the data are determined by the errors of the line parameters including the uncertainty of the intensity calibration (20\%).
The total (ortho plus para) column density is derived to be $(1.08\pm0.16)\times10^{14}$ cm$^{-2}$.
The ortho-to-para ratio is evaluated to be $3.1\pm0.4$, which is comparable to the statistical weight of 3. This ratio is consistent with the previous report by \citet{takakuwa} ($2.5\pm0.5$).
The rotation temperature is ($8.3\pm0.3$) K, which is lower than that reported for C$_4$H$_2$ in L1527 of 12.3 K \citep{s08}. The errors denote the standard deviation of the fit.

We also estimated the column densities of the two $^{13}$C species, c-$\mathrm{^{13}CCCH_2}$ and c-$\mathrm{CC^{13}CH_2}$, by assuming LTE condition.
The column density and the rotation temperature of c-$\mathrm{^{13}CCCH_2}$ are evaluated to be $(3.5\pm0.7)\times10^{11}$ cm$^{-2}$ and $(8.6\pm1.5)$ K, respectively, by assuming the same ortho-to-para ratio as the normal species.
On the other hand, the column density and the excitation temperature of c-$\mathrm{CC^{13}CH_2}$ are determined to be $(1.79\pm0.19)\times10^{12}$ cm$^{-2}$ and $(7.3\pm0.4)$ K, respectively.
The abundance ratio of [c-C$_3$H$_2$]/[c-$\mathrm{^{13}CCCH_2}$] is thus determined to be $310\pm80$, whereas that of [c-C$_3$H$_2$]/[c-$\mathrm{CC^{13}CH_2}$] is $61\pm11$. 
[c-$\mathrm{CC^{13}CH_2}$] represents the total abundance of the c-$\mathrm{CC^{13}CH_2}$ species without any correction for the statistical factor of 2 (Figure \ref{structure}; Section 3.1). These results are higher than the ratios expected for the condition that $^{13}$C is randomly distributed among the three carbon atoms and the elemental abundance ratio of $^{12}$C/$^{13}$C is 60--70, as shown in Table \ref{ratios}.

\begin{figure}
\begin{center}
\includegraphics[width=\textwidth]{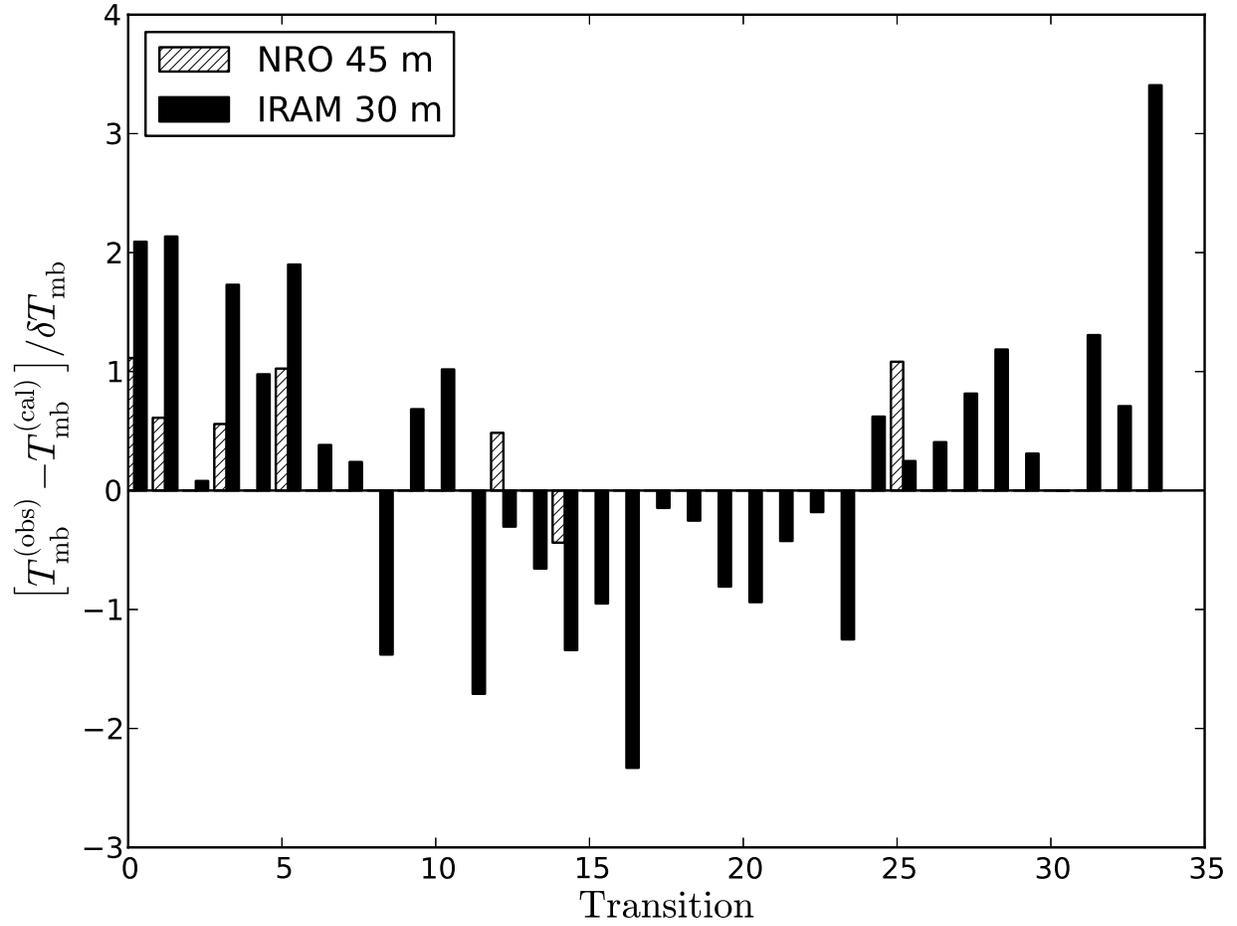}
\caption{Residuals of the LTE fit divided by uncertainty of each line, arranged in ascending order of the upper-state energy.}
\label{cmp1}
\end{center}
\end{figure}

Figure \ref{cmp1} shows the residuals in the least-squares fit divided by the uncertainty $\delta T$ of each line, $(T_\mathrm{obs}-T_\mathrm{fit})/\delta T$, where $T_\mathrm{obs}$ and $T_\mathrm{fit}$ are the observed intensity and the best-fit value, respectively. The uncertainty of the intensity $\delta T$ for each line is estimated by taking into account the Gaussian fitting error of the line profile and the calibration uncertainty.
Figure \ref{cmp1} shows a slight systematic residuals in the fit. Intensities of the lines with high and low upper state energies systematically show positive residuals, while those of the lines with moderate upper state energies tend to have negative residuals. Such a systematic trend can also be seen, even if the different source sizes of 15\arcsec\ and 30\arcsec\ are assumed. This may indicate that the source includes two components having different rotation temperatures. The physical structure of the source (i.e. the gradient in the density and the temperature) may also play a role. However, the systematic residuals are eliminated by a non-LTE analysis described below, and we therefore consider that the systematic residuals mainly originate from imperfection of the LTE assumption.

\subsection{LVG analysis for c-C$_3$H$_2$}
\begin{figure}
\begin{center}
\includegraphics[width=\textwidth]{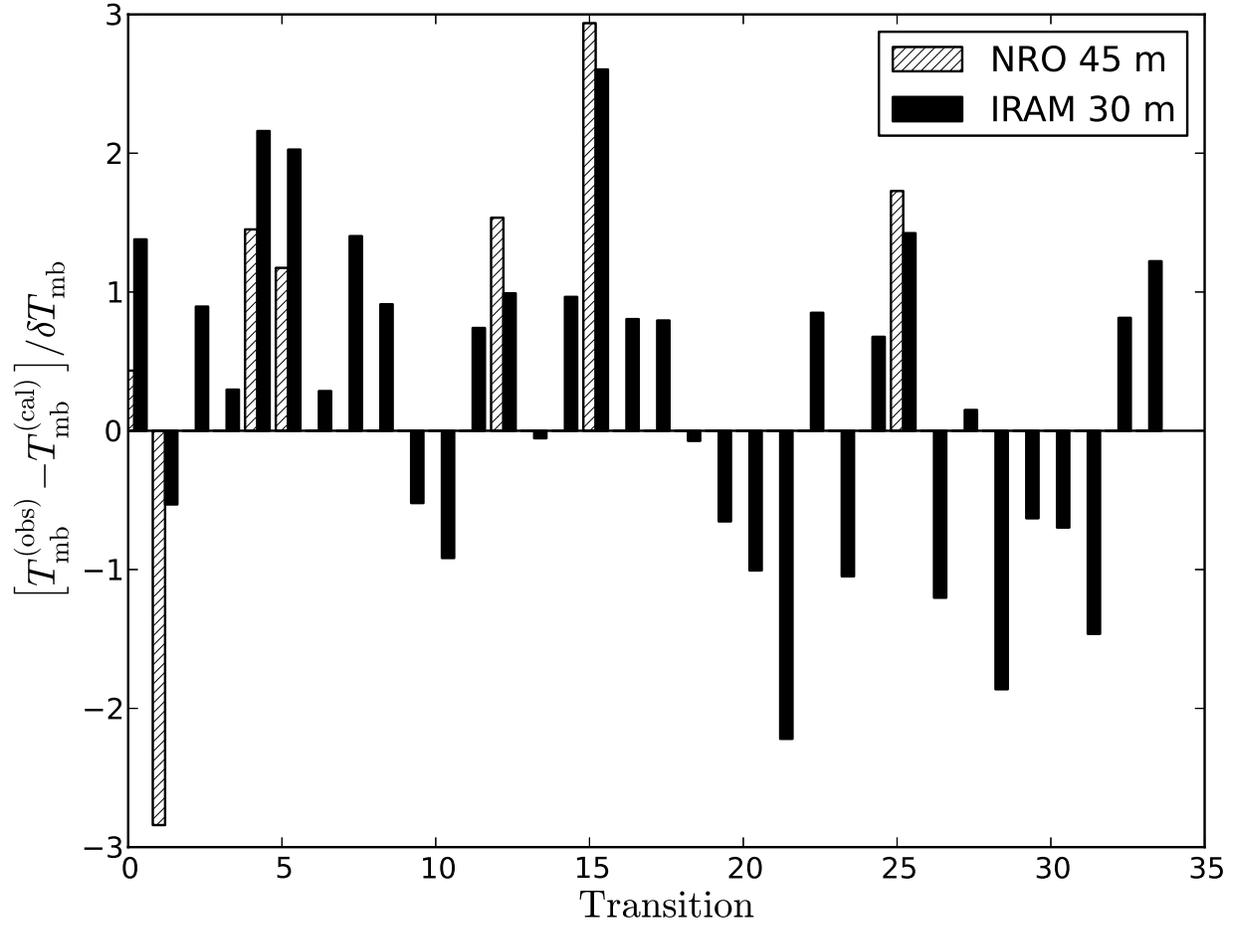}
\caption{The same as Figure \ref{cmp1}, but obtained by assuming the LVG condition.}
\label{cmp2}
\end{center}
\end{figure}

We employ a large velocity gradient (LVG) analysis based
on the non-LTE radiative transfer code Radex \citep{Tak} for the analysis of the normal species.
If the gas kinetic temperature is assumed to be 25 K \citep{s10b}, the H$_2$ density and the total column density of c-C$_3$H$_2$ are determined to be $(2.9\pm0.5)\times10^5$ cm$^{-3}$ and $(8.8\pm1.1)\times10^{13}$ cm$^{-2}$, respectively. Even if the gas kinetic temperature is assumed to be 10 K for reference, the total column density of c-C$_3$H$_2$ does not change very much ($(1.18\pm0.11)\times10^{14}$ cm$^{-2}$).
Figure \ref{cmp2} shows the residuals of the fit for the gas kinetic temperature of 25 K, calculated in the same way as in Figure \ref{cmp1}.
The systematic errors seen in Figure \ref{cmp1} are now eliminated in Figure \ref{cmp2}.
Although the non-LTE analysis is favorable as seen here, the collisional cross sections are not reported for the $^{13}$C isotopic species. 
Hence, we cannot determine the abundance ratios of the $^{13}$C isotopic species relative to the normal species by using the non-LTE analysis. 
Since the total column density derived from the non-LTE analysis only differs from that derived from the LTE analysis by 20\%, we employ the LTE results for discussions of the $^{12}$C/$^{13}$C ratio of c-C$_3$H$_2$.

\section{Discussion}
\subsection{Dilution of the $^{13}$C species}
The $^{12}$C/$^{13}$C ratios of c-C$_3$H$_2$ derived from the column densities are summarized in Table \ref{ratios}.
The averaged $^{12}$C/$^{13}$C ratio of c-C$_3$H$_2$ is calculated as 
\begin{equation}
R_\mathrm{av} = \frac{3[\mathrm{c\mathchar`-C_3H_2}]}{[\mathrm{c\mathchar`-^{13}CCCH_2}]+[\mathrm{c\mathchar`-CC^{13}CH_2}]}.
\end{equation}
The numerical factor 3 is introduced because three carbon atoms are available for the $^{13}$C substitution. 
The $R_\mathrm{av}$ values are evaluated to be $150\pm30$ under the LTE condition.
As shown in Table \ref{ratios}, the $^{12}$C/$^{13}$C ratios of c-C$_3$H$_2$ are thus found to be higher than those expected from the interstellar elemental abundance ratio of 60--70 \citep{lucas, milam}. For comparison, we have also derived the $^{12}$C/$^{13}$C ratio in L1527 by using the $J=1$--$0$ line data of C$^{18}$O and $^{13}$C$^{18}$O obtained in the ASAI and Nobeyama observations to be $68\pm19$. The $R_\mathrm{av}$ value obtained above is also higher than this ratio.

In dense molecular clouds, the main reservoir of $^{13}$C is $^{13}$CO, which is the most abundant carbon-bearing molecule.
As carbon-chain molecules are produced by reactions starting from C$^{+}$, their $^{13}$C species are formed by the reaction from $^{13}$C$^{+}$.  Although the $^{13}$C$^{+}$ ion is produced from $^{13}$CO with the reactions of He$^+$, as ${}^{12}$C$^{+}$ from ${}^{12}$CO, $^{13}$C$^{+}$ will go back to $^{13}$CO by the following isotope exchange reaction \citep{langer, roueff}:
\begin{equation}
\mathrm{^{13}C^{+}+{}^{12}CO\rightarrow {}^{13}CO+{}^{12}C^{+}+35\ K.}\label{exchange}
\end{equation}
This is the most important reaction for $^{13}$C$^{+}$, because the reaction with H$_2$ is a very slow radiative association. The electron recombination is also much slower than the reaction with $^{12}$CO.
This exchange reaction is exothermic, and hence, its backward reaction is slow in cold clouds. 
Then, the $^{12}$C$^{+}$/$^{13}$C$^{+}$ ratio tends to be higher in cold clouds.
The $^{13}$C species of various molecules produced from $^{13}$C$^{+}$ thus become less abundant. This mechanism was initially proposed by \citet{langer}. \citet{woods} and \citet{furuya} confirmed this trend by chemical model calculations. It should be noted that the isotope selective photodissociation, which could cause the $^{12}$C/$^{13}$C isotope anomaly in photodissociation regions, is not important in our observations, because we are observing a dense part near the protostar.

Such a dilution process can occur even in the protostellar envelope of L1527, whose gas kinetic temperature is 25 K \citep{s10b, s14}.
In order to verify this, we consider the following relation under the steady-state approximation:
\begin{equation}
\frac{[\mathrm{^{13}C^{+}}]}{[\mathrm{^{12}C^{+}}]}=\frac{k_\mathrm{b}[\mathrm{^{13}CO}]}{k_\mathrm{f}[\mathrm{{}^{12}CO}]},
\end{equation}
where $k_\mathrm{f}$ and $k_\mathrm{b}$ are the rate coefficients of the forward and backward reactions of (\ref{exchange}), respectively, and $[X]$ denotes the abundance of $X$.
These rate coefficients fulfill the following relation: 
\begin{equation}
k_\mathrm{b}=k_\mathrm{f}\exp\left(-\frac{\Delta G}{kT}\right),
\end{equation}
 where $\Delta G$ represents the free energy difference between the right and left hand sides of the reaction (\ref{exchange}). 
Since $k_\mathrm{b}/k_\mathrm{f}$ is evaluated to be 0.2 at 25 K, the dilution process of $^{13}$C$^{+}$ is barely possible in the warm region of L1527.

L1527 is a warm-carbon-chain-chemistry source, where CH$_4$ evaporated from dust grains in the warm ($\gtrsim25$ K) and the dense part near the protostar triggers efficient production of carbon-chain molecules including c-C$_3$H$_2$.
CH$_4$ is thought to be formed by hydrogenation from the neutral carbon atoms depleted onto dust grains in the cold starless core phase and/or the less dense phase. The $^{12}$C/$^{13}$C ratio of the carbon atom is also expected to be larger than the elemental ratio as the $^{12}$C$^+$/$^{13}$C$^+$ ratio, because the neutral carbon atom (C) is formed from electron recombination of the carbon ion (C$^+$). Hence, the $^{12}$C/$^{13}$C ratio of CH$_4$ formed on dust grains would also be higher.
This may also contribute to the dilution of the $^{13}$C species of c-C$_3$H$_2$.

\subsection{Abundance difference between two $^{13}$C species}
As described in the Introduction, it is reported that several carbon-chain molecules show the abundance anomaly of the $^{13}$C species. 
For CCH, CCS, C$_3$S, C$_4$H, and HC$_3$N, the $^{13}$C abundances for different carbon atoms in a single molecular species are different from one another.
In order to compare the abundance of the two $^{13}$C species (Section 3.1), we calculate 
the $[\mathrm{c\mathchar`-^{13}CCCH_2}]/[\mathrm{c\mathchar`-CC^{13}CH_2}]$ ratio to be $0.20\pm0.05$. If $^{13}$C were randomly substituted for the three carbon atoms, the [c-$\mathrm{^{13}CCCH_2}$]/[c-$\mathrm{CC^{13}CH_2}$] ratio would be 0.5. Hence, the observed ratio indicates that c-$\mathrm{CC^{13}CH_2}$ exists more favorably than c-$\mathrm{^{13}CCCH_2}$.
It has been thought that the main production pathway for c-C$_3$H$_2$ is the electron recombination of $\mathrm{C_3H_3^+}$.
This is because the cyclopropenyl cation $\mathrm{C_3H_3^+}$ is a major ionic species, which is produced by the gas-phase reactions starting from CH$_4$ \citep{sy13}.
Since $\mathrm{C_3H_3^+}$ has three equivalent carbon atoms, c-C$_3$H$_2$ produced from $\mathrm{C_3H_3^+}$ cannot contribute to the abundance anomaly of $^{13}$C species.
Hence, other reactions have to be considered as the dominant production pathway (e.g., $\mathrm{C_2H_2+CH\rightarrow C_3H_2+H}$). 

Another possible reason for the abundance anomaly of the two $^{13}$C species is the exchange reaction between c-$\mathrm{^{13}CCCH_2}$ and c-$\mathrm{CC^{13}CH_2}$, although it is uncertain whether this reaction happens for the closed shell molecule, c-C$_3$H$_2$.
The zero-point vibrational energies of c-$\mathrm{^{13}CCCH_2}$ and c-$\mathrm{CC^{13}CH_2}$ are evaluated to be lower than the normal species by 8.74 K and 53.2 K, respectively, by using the molecular constants by \citet{dateo}. Hence, it would be possible that c-$\mathrm{CC^{13}CH_2}$ is enhanced relative to c-$\mathrm{^{13}CCCH_2}$, if the exchange reaction is possible. In case of $^{13}$CCH and C${}^{13}$CH, the reaction such as 
\begin{equation}
\mathrm{^{13}CCH+H\rightarrow C{}^{13}CH+H}
\end{equation}
is proposed \citep{s10a}. It is thus interesting to explore whether a similar process is possible for c-C$_3$H$_2$.
It should be noted that the isotopologue with the lowest zero-point vibrational energy tends to have higher abundances for CCH, CCS, and C$_3$S, although this trend is uncertain for C$_4$H \citep{s07,s10a,s13}. This fact may suggest possible contribution of the position exchange reactions \citep{s10a,furuya}. More detailed studies of these reactions as well as chemical model calculations for the $^{13}$C species in molecular clouds are awaited.

\section{Concluding Remarks}
In this study, we accurately determined the [c-C$_3$H$_2$]/[c-$\mathrm{^{13}CCCH_2}$] and [c-C$_3$H$_2$]/[c-$\mathrm{CC^{13}CH_2}$] ratios on the basis of the extensive data of the Nobeyama and ASAI observations. We found a significant dilution of the $^{13}$C species, and a significant difference between the [c-C$_3$H$_2$]/[c-$\mathrm{^{13}CCCH_2}$] and [c-C$_3$H$_2$]/[c-$\mathrm{CC^{13}CH_2}$] ratios, even when the statistical factor of 2 for the c-$\mathrm{CC^{13}CH_2}$ species is taken into account. This further confirms that the $^{13}$C anomaly is an important phenomenon in astrochemistry.

It should be noted that determination of the $^{12}$C/$^{13}$C ratio requires accurate determination of the column density of the normal species. This was indeed possible in this study, where a large number of lines can be used to conduct a fine analysis. However, this is not always the case. On the other hand, the anomaly in the $^{12}$C/$^{13}$C ratio among the different carbon atoms in a single molecule can be studied only by observations of the $^{13}$C species, which often emit optically thin lines. If the cause of the $^{12}$C/$^{13}$C anomaly in a single molecule is established, it could be a new tracer to understand chemical processes and physical conditions of molecular clouds. 

The anomalies of the $^{13}$C species found for c-C$_3$H$_2$ and other carbon-chain molecules might be related to the $^{13}$C anomaly in meteorites \citep{floss}. 
Although the $^{12}$C/$^{13}$C anomaly in dust forming regions of AGB stars \citep[e.g.,][]{milam09, kahane, tomkin} mainly contributes to the anomaly in meteorites, the $^{12}$C/$^{13}$C anomaly in cold molecular clouds may also be responsible.
Hence, understanding the abundance anomaly of the $^{13}$C species and its evolution in star-forming regions is of particular importance to trace back from our planetary system to its interstellar origins.

\acknowledgments
This study is supported by Grant-in-Aids from Ministry of Education, Culture, Sports, Science, and Technologies of Japan (21224002, 25400223 and 25108005). 
The authors also acknowledge financial support by JSPS and MAEE under the Japan-France integrated action program (SAKURA:25765VC).

\newpage
\clearpage
\begin{deluxetable}{ccccccccccccc}
\tabletypesize{\tiny}
\rotate
\tablecaption{Observed line parameters of c-C$_3$H$_2$ toward L1527\label{normal}}
\tablewidth{0pt}
\tablehead{
\colhead{Frequency}					& \colhead{Transition}       				& 
\colhead{o/p}					&
\colhead{$E_\mathrm{u}$} 				& \colhead{$\langle S\mu^2\rangle$} 		& 	
\colhead{$\Delta V_\mathrm{LSR}$\tablenotemark{a}} 			& \colhead{$\Delta v$\tablenotemark{a}}           				& 
\colhead{rms}					  		& \colhead{$T_\mathrm{mb}$\tablenotemark{a}} &
\colhead{$\int T_\mathrm{mb}dv$} &
\colhead{$\sigma_\mathrm{LTE}$\tablenotemark{b}} 	& 
\colhead{$\sigma_\mathrm{LVG}$\tablenotemark{c}} 		& 
\colhead{telescope} \\
(GHz)			& 				&	
&
(K)		& (Debye$^2$)		&
(km s$^{-1}$) 	& (km s$^{-1}$) 	&
(mK) 			& (K) 			&
(K km s$^{-1}$) 	&  				& 
				&
	}
\startdata
80.723180&$4_{2,2}$--$4_{1,3}$&p&28.8&19.5&$-0.07(4)$&1.72(9)&5.8&0.12(3)&$0.25(5)$&$0.48$&$1.54$&NRO\\
&&&&&                                     $-0.09(6)$&0.81(10)&5.9&0.14(3)&$0.11(2)$&$-0.30$&$0.99$&IRAM\\
82.093544&$2_{0,2}$--$1_{1,1}$&p&6.4&14.6&$-0.15(13)$&1.40(3)&34.3&1.5(3)&$2.5(5)$&$1.11$&$0.43$&NRO\\
&&&&&                                     $-0.030(4)$&0.820(11)&3.9&1.8(4)&$1.6(3)$&$2.09$&$1.38$&IRAM\\
82.966200&$3_{1,2}$--$3_{0,3}$&o&16.0&31.3&$-0.101(7)$&1.255(17)&15.2&1.1(2)&$1.6(3)$&$0.56$&$1.45$&NRO\\
&&&&&                                      $-0.063(4)$&0.788(5)&3.5&1.4(3)&$1.2(2)$&$1.73$&$2.16$&IRAM\\
84.727688&$3_{2,2}$--$3_{1,3}$&p&16.1&10.3&$0.150(5)$&1.154(12)&5.5&0.52(11)&$0.67(13)$&$1.02$&$1.17$&NRO\\
&&&&&                                      $-0.037(5)$&1.196(10)&2.7&0.42(8)&$0.54(11)$&$1.90$&$2.03$&IRAM\\
85.338894&$2_{1,2}$--$1_{0,1}$&o&6.4&48.1&$0.340(6)$&1.239(14)&21.1&2.8(6)&$3.8(8)$&$0.61$&$-2.84$&NRO\\
&&&&&                                      $-0.070(5)$&1.065(16)&2.9&2.8(6)&$3.3(7)$&$2.14$&$-0.53$&IRAM\\
85.656431&$4_{3,2}$--$4_{2,3}$&o&29.1&56.2&$-0.010(9)$&1.16(2)&7.4&0.46(9)&$0.57(11)$&$-0.44$&$2.94$&NRO\\
&&&&&                                      $-0.060(9)$&1.15(3)&2.1&0.26(5)&$0.33(7)$&$-1.34$&$2.60$&IRAM\\
87.435318&$5_{4,2}$--$5_{3,3}$&p&45.3&26.2&$-0.2(2)$&2.0(5)&10.5&0.032(9)&$0.045(19)$&$1.08$&$1.73$&NRO\\
&&&&&                                      $0.00(11)$&1.9(3)&1.8&0.016(4)&$0.036(8)$&$0.25$&$1.42$&IRAM\\
90.344082&$6_{5,2}$--$6_{4,3}$&o&64.7&98.3&$-0.16(19)$&1.2(5)&3.8&0.014(5)&$0.017(7)$&$0.71$&$0.81$&IRAM\\
145.089606&$3_{1,2}$--$2_{2,1}$&o&16.0&40.0&$0.004(2)$&0.727(5)&12.8&3.0(6)&$2.3(5)$&$0.98$&$0.30$&IRAM\\
150.436555&$2_{2,0}$--$1_{1,1}$&p&9.7&6.8&$-0.136(2)$&0.634(8)&8.8&1.7(3)&$1.2(2)$&$0.080$&$0.89$&IRAM\\
150.820665&$4_{0,4}$--$3_{1,3}$&p&19.3&36.9&$-0.162(4)$&0.779(10)&8.2&2.3(5)&$1.9(4)$&$0.68$&$-0.52$&IRAM\\
150.851908&$4_{1,4}$--$3_{0,3}$&o&19.3&110.7&$-0.201(13)$&0.92(3)&8.3&3.3(7)&$3.2(6)$&$1.02$&$-0.92$&IRAM\\
150.954691&$6_{2,4}$--$6_{1,5}$&p&54.7&20.3&$-0.23(8)$&0.6(3)&9.3&0.05(2)&$0.025(9)$&$0.31$&$-0.63$&IRAM\\
151.039173&$6_{3,4}$--$6_{2,5}$&o&54.7&60.8&$-0.10(4)$&0.71(9)&8.8&0.10(2)&$0.085(20)$&$-0.010$&$-0.70$&IRAM\\
151.343878&$5_{1,4}$--$5_{0,5}$&o&35.4&32.4&$-0.065(8)$&0.811(17)&6.7&0.43(9)&$0.37(8)$&$-0.15$&$0.79$&IRAM\\
151.361105&$5_{2,4}$--$5_{1,5}$&p&35.4&10.8&$0.00(2)$&0.75(5)&6.4&0.14(3)&$0.11(2)$&$-0.94$&$-0.65$&IRAM\\
155.518303&$3_{2,2}$--$2_{1,1}$&p&16.1&17.8&$0.005(2)$&0.687(5)&8.2&2.0(4)&$1.5(3)$&$0.38$&$0.29$&IRAM\\
204.788926&$4_{2,2}$--$3_{3,1}$&p&28.8&11.2&$-0.036(7)$&0.761(16)&7.4&0.52(10)&$0.44(9)$&$-0.66$&$-0.054$&IRAM\\
216.278756&$3_{3,0}$--$2_{2,1}$&o&19.5&45.6&$-0.128(7)$&0.892(15)&5.2&2.0(4)&$1.9(4)$&$-1.71$&$0.74$&IRAM\\
217.822148&$6_{0,6}$--$5_{1,5}$, $6_{1,6}$--$5_{0,5}$&p/o&38.6&58.3&$-0.092(3)$&0.922(8)&4.4&2.5(5)&$2.5(5)$&$-0.42$&$-2.22$&IRAM\\
217.940046&$5_{1,4}$--$4_{2,3}$&o&35.4&110.3&$-0.057(2)$&0.863(5)&5.5&1.8(4)&$1.7(3)$&$-0.25$&$-0.073$&IRAM\\
218.160456&$5_{2,4}$--$4_{1,3}$&p&35.4&36.8&$-0.022(4)$&0.815(11)&6.5&0.73(15)&$0.63(13)$&$-0.81$&$-1.01$&IRAM\\
218.732732&$7_{2,6}$--$7_{1,7}$, $7_{1,6}$--$7_{0,7}$&p/o&61.2&11.0&$0.05(3)$&0.92(8)&4.4&0.071(15)&$0.084(18)$&$1.31$&$-1.46$&IRAM\\
227.169138&$4_{3,2}$--$3_{2,1}$&o&29.1&61.6&$-0.093(4)$&0.827(9)&8.4&1.8(4)&$1.6(3)$&$-0.95$&$0.96$&IRAM\\
244.222150&$3_{2,1}$--$2_{1,2}$&o&18.2&7.3&$-0.047(3)$&0.761(6)&6.0&1.5(3)&$1.2(2)$&$0.24$&$1.40$&IRAM\\
249.054368&$5_{2,3}$--$4_{3,2}$&o&41.0&76.3&$-0.050(4)$&0.791(10)&5.1&1.0(2)&$0.89(18)$&$-0.18$&$0.85$&IRAM\\
251.314367&$7_{1,7}$--$6_{0,6}$, $7_{0,7}$--$6_{1,6}$&p/o&50.7&69.0&$-0.014(3)$&0.830(8)&5.7&1.6(3)&$1.5(3)$&$1.19$&$-1.86$&IRAM\\
251.508709&$6_{1,5}$--$5_{2,4}$&p&47.5&47.4&$0.029(9)$&0.74(2)&6.0&0.41(8)&$0.35(7)$&$0.41$&$-1.20$&IRAM\\
251.527311&$6_{2,5}$--$5_{1,4}$&o&47.5&142.2&$-0.005(4)$&0.791(10)&6.4&1.1(2)&$1.0(2)$&$0.81$&$0.15$&IRAM\\
252.409837&$8_{1,7}$--$8_{0,8}$, $8_{2,7}$--$8_{1,8}$&p/o&76.5&11.1&$-0.05(6)$&0.94(14)&5.3&0.051(12)&$0.038(10)$&$3.41$&$1.22$&IRAM\\
254.987652&$5_{3,3}$--$4_{2,2}$&p&41.1&26.8&$0.048(8)$&0.752(18)&5.6&0.36(7)&$0.30(6)$&$-1.25$&$-1.04$&IRAM\\
260.479746&$5_{3,2}$--$4_{4,1}$&o&44.7&25.8&$0.00(11)$&0.83(3)&6.8&0.32(6)&$0.31(6)$&$0.62$&$0.68$&IRAM\\
261.831811&$3_{3,1}$--$2_{0,2}$&p&19.0&1.4&$0.02(3)$&0.68(6)&18.4&0.30(6)&$0.19(4)$&$-1.38$&$0.91$&IRAM\\
265.759481&$4_{4,1}$--$3_{3,0}$&o&32.2&89.8&$-0.076(5)$&0.879(13)&8.2&1.4(3)&$1.4(3)$&$-2.33$&$0.80$&IRAM\\
\enddata
\tablecomments{The numbers in parentheses represent the errors in unit of last significant digits.}
\tablenotetext{a}{Obtained by the Gaussian fit.}
\tablenotetext{b}{The residual divided by the intensity error, shown in Figure \ref{cmp1}.}
\tablenotetext{c}{The same as $\sigma_\mathrm{LTE}$, but shown in Figure \ref{cmp2}.}
\end{deluxetable}
\clearpage
\begin{deluxetable}{ccccccccccc}
\tabletypesize{\tiny}
\rotate
\tablecaption{Observed line parameters of c-CC$^{13}$CH$_2$ toward L1527\label{off}}
\tablewidth{0pt}
\tablehead{
\colhead{Frequency}					& \colhead{Transition}       				& 
\colhead{$E_\mathrm{u}$} 				& \colhead{$\langle S\mu^2\rangle$} 		& 	
\colhead{$\Delta V_\mathrm{LSR}$\tablenotemark{a}} 			& \colhead{$\Delta v$\tablenotemark{a}}           				& 
\colhead{rms}					  		& \colhead{$T_\mathrm{mb}$\tablenotemark{a}} &
\colhead{$\int T_\mathrm{mb}dv$} & 
\colhead{$\sigma_\mathrm{LTE}$\tablenotemark{b}} 	& \colhead{telescope} \\
(GHz)			& 				&	
(K)		& (Debye$^2$)		&
(km s$^{-1}$) 	& (km s$^{-1}$) 	&
(mK) 			& (K) 			&
(K km s$^{-1}$) 	&
  				&  				 }
\startdata
80.047537&$2_{0,2}$--$1_{1,1}$&6.3&27.8&$-0.07(2)$&1.25(5)&7.5&0.16(3)&$0.23(5)$&$1.68$&NRO\\
80.775347&$3_{1,2}$--$3_{0,3}$&15.7&21.0&$0.19(16)$&1.5(4)&6.4&0.021(6)&$0.036(11)$&$0.29$&NRO\\
83.474137&$3_{2,2}$--$3_{1,3}$&15.9&20.5&$-0.00(16)$&1.5(4)&6.8&0.022(7)&$0.033(12)$&$0.35$&NRO\\
&&&&                                     $0.00(2)$&1.1(5)&3.5&0.018(5)&$0.027(7)$&$0.053$&IRAM\\
84.185635&$2_{1,2}$--$1_{0,1}$&6.3&31.3&$0.015(10)$&1.18(2)&4.3&0.21(4)&$0.28(6)$&$1.65$&NRO\\
&&&&                                     $-0.08(2)$&0.78(5)&3.6&0.26(5)&$0.22(4)$&$2.29$&IRAM\\
114.897371&$3_{0,3}$--$2_{1,2}$&11.8&50.5&$0.00(4)$&0.90(6)&19.6&0.32(7)&$0.30(6)$&$0.80$&NRO\\
115.524356&$3_{1,3}$--$2_{0,2}$&11.8&50.7&$-0.00(6)$&0.78(7)&22.3&0.31(7)&$0.32(7)$&$-0.052$&NRO\\
140.432881&$3_{1,2}$--$2_{2,1}$&15.7&23.9&$-0.13(3)$&0.53(10)&7.3&0.15(4)&$0.081(18)$&$0.25$&IRAM\\
147.702239&$2_{2,0}$--$1_{1,1}$&9.6&13.9&$-0.18(3)$&0.65(5)&9.4&0.17(4)&$0.10(2)$&$0.021$&IRAM\\
148.114191&$4_{1,4}$--$3_{0,3}$&19.0&71.6&$-0.161(18)$&0.78(4)&6.9&0.18(4)&$0.14(3)$&$-0.40$&IRAM\\
153.894626&$3_{2,2}$--$2_{1,1}$&15.9&34.8&$-0.10(2)$&0.63(6)&9.2&0.16(4)&$0.13(3)$&$-0.83$&IRAM\\
212.457672&$3_{3,0}$--$2_{2,1}$&19.2&31.7&$-0.02(5)$&0.94(11)&7.3&0.081(18)&$0.083(19)$&$-3.06$&IRAM\\
213.843018&$6_{1,6}$--$5_{0,5}$, $6_{0,6}$--$5_{1,5}$&37.9&113.5&$0.02(3)$&0.96(6)&4.4&0.088(18)&$0.085(18)$&$0.66$&IRAM\\
213.872779&$5_{1,4}$--$4_{2,3}$&34.8&70.7&$0.06(5)$&0.79(11)&4.4&0.045(10)&$0.029(7)$&$0.11$&IRAM\\
214.313034&$5_{2,4}$--$4_{1,3}$&34.8&70.8&$-0.07(5)$&0.82(11)&4.2&0.052(12)&$0.060(13)$&$0.78$&IRAM\\
225.435934&$4_{3,2}$--$3_{2,1}$&28.6&40.3&$0.02(9)$&1.00(2)&7.9&0.051(14)&$0.044(14)$&$-0.027$&IRAM\\
237.998109&$3_{2,1}$--$2_{1,2}$&17.8&5.1&$-0.08(6)$&0.71(14)&4.3&0.037(10)&$0.019(6)$&$-0.53$&IRAM\\
246.723236&$7_{1,7}$--$6_{0,6}$, $7_{0,7}$--$6_{1,6}$&49.7&134.3&$0.21(6)$&0.70(14)&5.6&0.044(12)&$0.040(10)$&$1.25$&IRAM\\
\enddata
\tablecomments{The numbers in parentheses represent the errors in unit of last significant digits.}
\tablenotetext{a}{Obtained by the Gaussian fit.}
\tablenotetext{b}{Residuals of the LTE fit divided by the uncertainty of each line.}
\end{deluxetable}
\clearpage
\begin{deluxetable}{cccccccccccc}
\tabletypesize{\tiny}
\rotate
\tablecaption{Observed line parameters of c-$^{13}$CCCH$_2$ toward L1527\label{on}}
\tablewidth{0pt}
\tablehead{
\colhead{Frequency}					& \colhead{Transition}       				& 
\colhead{o/p}					&
\colhead{$E_\mathrm{u}$} 				& \colhead{$\langle S\mu^2\rangle$} 		& 	
\colhead{$\Delta V_\mathrm{LSR}$\tablenotemark{a}} 			& \colhead{$\Delta v$\tablenotemark{a}}           				& 
\colhead{rms}					  		& \colhead{$T_\mathrm{mb}$\tablenotemark{a}} &
\colhead{$\int T_\mathrm{mb}dv$} &
\colhead{$\sigma_\mathrm{LTE}$\tablenotemark{b}} 	& \colhead{telescope} \\
(GHz)			& 				&	
&
(K)		& (Debye$^2$)		&
(km s$^{-1}$) 	& (km s$^{-1}$) 	&
(mK) 			& (K) 			&
(K km s$^{-1}$) 	&
  				&  				 }
\startdata
81.150881&$2_{0,2}$--$1_{1,1}$&p&6.3&31.0&$0.10(18)$&1.3(4)&5.5&0.016(5)&$0.021(8)$&$0.84$&NRO\\
&&&&&$-0.00(3)$&0.9(3)&4.1&0.018(6)&$0.026(8)$&$1.17$&IRAM\\
82.303747&$2_{1,2}$--$1_{0,1}$&o&6.3&96.2&$-0.17(12)$&1.1(3)&12.6&0.048(14)&$0.049(16)$&$0.22$&NRO\\
&&&&&$-0.08(7)$&0.83(8)&3.9&0.064(14)&$0.060(13)$&$2.07$&IRAM\\
114.381212&$3_{0,3}$--$2_{1,2}$&o&11.8&159.0&$-0.15(11)$&1.1(2)&14.8&0.08(2)&$0.07(3)$&$0.30$&NRO\\
145.353129&$3_{1,2}$--$2_{2,1}$&o&15.7&96.1&$0.10(11)$&0.6(2)&10.3&0.044(16)&$0.033(13)$&$-0.80$&IRAM\\
211.613800&$3_{3,0}$--$2_{2,1}$&o&18.9&75.0&$0.02(7)$&0.69(17)&4.9&0.033(10)&$0.024(7)$&$-2.15$&IRAM\\
212.387110&$6_{0,6}$--$5_{1,5}$, $6_{1,6}$--$5_{0,5}$&p/o&37.6&117.2&$0.18(9)$&0.8(2)&6.5&0.039(12)&$0.026(8)$&$0.39$&IRAM\\
245.044421&$7_{1,7}$--$6_{0,6}$, $7_{0,7}$--$6_{1,6}$&p/o&49.4&138.6&$-0.04(10)$&1.2(2)&4.8&0.029(8)&$0.033(10)$&$2.72$&IRAM\\
\enddata
\tablecomments{The numbers in parentheses represent the errors in unit of last significant digits.}
\tablenotetext{a}{Obtained by the Gaussian fit.}
\tablenotetext{b}{Residuals of the LTE fit divided by the uncertainty of each line.}
\end{deluxetable}
\clearpage
\begin{table}
\begin{center}
\caption{Number of detected lines.\label{N_of_lines}}
\begin{tabular}{cccccc}
\tableline\tableline
 & NRO 45 m &\multicolumn{3}{c}{IRAM 30 m} & \\
\cline{2-2}\cline{3-5}
Species & 3 mm & 3 mm & 2 mm & 1 mm& Total \\
\tableline
c-C$_3$H$_2$&7\tablenotemark{a}&8&9&17&41\\
c-CC$^{13}$CH$_2$&6&2\tablenotemark{b}&4&7&19\\
c-$^{13}$CCCH$_2$&3&2\tablenotemark{b}&1&3&9\\
\tableline
\tablenotetext{a}{All the transitions are also detected with IRAM 30 m.}
\tablenotetext{b}{All the transitions are also detected with NRO 45 m.}
\end{tabular}
\end{center}
\end{table}
\clearpage
\begin{table}
\begin{center}
\caption{The observed $^{12}$C/$^{13}$C ratios of c-C$_3$H$_2$. \label{ratios}}
\begin{tabular}{ccc}
\tableline\tableline
Ratio&Observed\tablenotemark{a}&Expected\tablenotemark{b}\\
\tableline
$\mathrm{[}$c-C$_3$H$_2\mathrm{]/[}$c-CC$^{13}$CH$_2$]&$61\pm11$&30--35\\
$\mathrm{[}$c-C$_3$H$_2\mathrm{]/[}$c-$^{13}$CCCH$_2$]&$310\pm80$&60--70\\
3[c-C$_3$H$_2\mathrm{]/([}$c-CC$^{13}$CH$_2$]+[c-$^{13}$CCCH$_2$])&$150\pm30$&60--70\\
$\mathrm{[}$c-$^{13}$CCCH$_2\mathrm{]/[}$c-CC$^{13}$CH$_2$]&$0.20\pm0.05$&0.5\\
\tableline
\tablecomments{The errors represent the standard deviation.}
\tablenotetext{a}{Ratios directly derived from the column densities of c-C$_3$H$_2$, c-$^{13}$CCCH$_2$, and c-CC$^{13}$CH$_2$.}
\tablenotetext{b}{Ratios expected for the condition that $^{13}$C is randomly distributed among the three carbon atoms and the elemental abundance ratio of $^{12}$C/$^{13}$C is 60--70.}

\end{tabular}
\end{center}
\end{table}
\end{document}